\documentclass[twocolumn,showpacs,prl]{revtex4}
\usepackage{amssymb}
\usepackage{graphicx}
\usepackage{dcolumn}
\usepackage{bm}
\usepackage{amsmath}

\begin{document}

\title{Long-lived qubit memory using atomic ions}
\date{April 10, 2005}

\author{C.~Langer}
\email{clanger@boulder.nist.gov}
\author{R.~Ozeri}
\author{J.~D.~Jost}
\author{J.~Chiaverini}
\author{B.~DeMarco}
\altaffiliation[Present address: ]{Department of Physics,
University of Illinois at Urbana-Champaign, Urbana, IL 61801-3080}
\author{A.~Ben-Kish}
\altaffiliation[Present address: ]{Department of Physics, Technion
- Israel Institute of Technology, Technion City, Haifa 32000,
Israel}
\author{R.~B.~Blakestad}
\author{J.~Britton}
\author{D.~B.~Hume}
\author{W.~M.~Itano}
\author{D.~Leibfried}
\author{R.~Reichle}
\author{T.~Rosenband}
\author{T.~Schaetz}
\altaffiliation[Present address: ]{Max Planck Inst Quantum Opt,
Hans Kopfermann Str 1, Garching, D-85748 Germany}
\author{P.~O.~Schmidt}
\altaffiliation[Present address: ]{Institute for Experimental
Physics, University of Innsbruck Technikerstr. 25, 5020 Innsbruck,
Austria}
\author{D.~J.~Wineland}
\affiliation{National Institute of Standards and Technology, 325
Broadway, Boulder, Colorado 80305, USA}

\begin{abstract}
We demonstrate experimentally a robust quantum memory using a
magnetic-field-independent hyperfine transition in $^{9}$Be$^{+}$
atomic ion qubits at a magnetic field $B \simeq 0.01194$~T. We
observe that the single physical qubit memory coherence time is
greater than 10 seconds, an improvement of approximately five
orders of magnitude from previous experiments with $^{9}$Be$^{+}$.
We also observe long coherence times of decoherence-free subspace
logical qubits comprising two entangled physical qubits and
discuss the merits of each type of qubit.
\end{abstract}


\maketitle

Scalable quantum information processing (QIP) requires physical
systems capable of reliably storing coherent superpositions for
periods over which quantum error correction can be
implemented~\cite{preskill}. Moreover, suppressing memory error
rates to very low levels allows for simpler error-correcting
algorithms~\cite{steane03,knill98}. In many current atomic ion QIP
experiments, a dominant source of memory error is decoherence
induced by fluctuating ambient magnetic
fields~\cite{barrett04,riebe04}. To address this problem, we
investigate creating long-lived qubit memories using a first-order
magnetic-field-independent hyperfine transition and logical qubits
of a decoherence-free subspace~\cite{DFS}.

Atomic systems have proven themselves as good candidates for
quantum information storage through their use in highly stable
atomic clocks~\cite{clocks}. Here, the principle of using
first-order magnetic-field-independent transitions is well
established. A typical clock transition $|F, m_{F} = 0\rangle
\leftrightarrow |F', m_{F'} = 0\rangle$ between hyperfine states
of angular momentum $F$ and $F'$ in alkali atoms has no linear
Zeeman shift at zero magnetic field, and coherence times exceeding
10 minutes have been observed~\cite{fisk}. Unfortunately, magnetic
sublevels in each hyperfine manifold are degenerate at zero
magnetic field. This makes it more advantageous to operate at a
nonzero field in order to spectrally resolve the levels, thereby
inducing a linear field dependence of the transition frequency.
However, field-independent transitions between hyperfine states
also exist at nonzero magnetic field.  In the context of atomic
clocks, coherence times exceeding 10 minutes have been observed in
$^{9}$Be$^{+}$ ions at a magnetic field $B =
0.8194$~T~\cite{bollinger}.

In neutral-atom systems suitable for QIP, field-independent
transitions at nonzero magnetic field have been investigated in
rubidium~\cite{harber02,treutlein04}. The radio-frequency
(RF)/microwave two-photon stimulated-Raman hyperfine transition
$|F=1,m_F=-1\rangle \leftrightarrow |F'=2,m_{F'}=1\rangle$ is
field-independent at approximately $3.23 \times 10^{-4}$~T , and
coherence times of 2.8~s have been observed~\cite{treutlein04}. In
these and the clock experiments, transitions were driven by
microwave fields on large numbers of atoms. Using microwaves, it
may be difficult to localize the fields well enough to drive
individual qubits unless a means (e.g., a magnetic-field gradient
or Stark-shift gradient) is employed to provide spectral
selection~\cite{mintert01,schrader04}, a technique that has the
additional overhead of keeping track of the phases induced by
these shifts. With transitions induced by laser beams, the
addressing can be accomplished by strong focusing~\cite{riebe04}
or by weaker focusing and inducing transitions in separate trap
zones~\cite{barrett04}. In contrast to microwave fields, optical
fields (using appropriate geometry~\cite{cirac95,sideband_cool})
provide stronger field gradients that are desirable for coupling
ion motional states with internal states, a requirement for
certain universal multi-qubit logic gates~\cite{cirac95,bible}.
Here, we explore the coherence time of a single atomic ion qubit
in a scalable QIP architecture using laser beam addressing.

\begin{figure}
\includegraphics{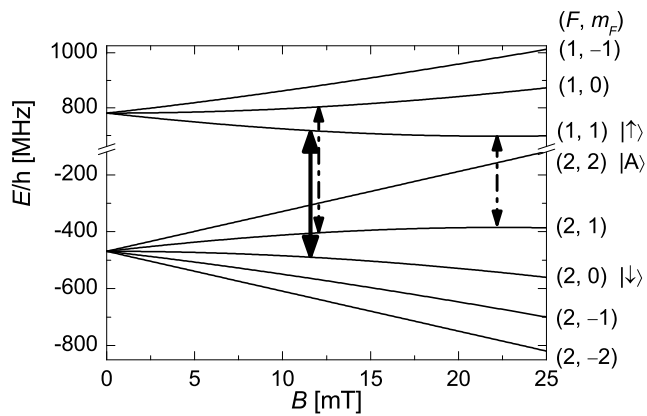} \caption{\label{breitrabi}Hyperfine level
structure of the $2s$~$^{2}S_{1/2}$ state of $^{9}$Be$^{+}$.  The
solid arrow indicates the magnetic-field-independent transition
studied here (0.01194~T); the dashed arrows indicate other useful
field-independent transitions at 0.01196~T and 0.02231~T.}
\end{figure}

In recent $^{9}$Be$^{+}$ QIP experiments utilizing 2s
$^{2}S_{1/2}$ hyperfine states: $|F = 2, m_{F} = -2\rangle$ and
$|F = 1, m_{F} = -1\rangle$ as qubit levels, fluctuating ambient
magnetic fields caused significant
decoherence~\cite{barrett04,ecorr}.  There, the qubit transition
depended linearly on the magnetic field with a coefficient of
approximately 21~kHz/$\mu$T (Fig.~\ref{breitrabi}).  Thus, random
magnetic field changes of 0.1~$\mu$T (typical in our laboratories)
would dephase qubit superpositions (to a phase uncertainty of 1
rad) in 80~$\mu$s. To mitigate this decoherence, refocussing
spin-echo $\pi$-pulses were inserted in the experimental
sequences~\cite{barrett04,ecorr} to limit the bandwidth of noise
to which the qubits were susceptible. However, these effects could
not be eliminated completely, and fluctuating fields remained a
major source of error in these experiments.

The energy spectrum of the ground hyperfine states of
$^{9}$Be$^{+}$ as a function of magnetic field is shown in
Fig.~\ref{breitrabi}.  At $B_{0} \simeq 0.01194$~T, the transition
$|F = 2, m_{F} = 0\rangle$~$\equiv$~$|$$\downarrow$$\rangle
\leftrightarrow |F = 1, m_{F} =
1\rangle$~$\equiv$~$|$$\uparrow$$\rangle$ (frequency
$\nu_{\uparrow\downarrow} \simeq$ 1.2~GHz) is first-order
field-independent with second-order dependence given by
$(0.305$~Hz$/\mu$T$^{2})(B-B_{0})^{2}$. Given random magnetic
field changes of 0.1~$\mu$T, we expect superpositions of
$|$$\downarrow\rangle$ and $|$$\uparrow\rangle$ to dephase in
approximately 50~s. The transition $|F = 2, m_{F} = 2\rangle
\equiv |A\rangle \leftrightarrow |$$\uparrow$$\rangle$ (frequency
$\nu_{\uparrow A} \simeq$ 1.0~GHz) is first-order field sensitive
with linear dependence of 17.6~kHz/$\mu$T for $B = B_0$.  We use
frequency measurements of this transition as a probe of the
magnetic field. We note that the $|F=2,m_F=1\rangle
\leftrightarrow |F'=1,m_{F'}=-1\rangle$ transition, similar to
that in Rb~\cite{harber02,treutlein04}, is field-independent in
$^9$Be$^+$ at $B = 2.54 \times 10^{-5}$~T; however, using detuned
laser excitation fields, this transition is less practical as it
is a four-photon transition.

In the experiment, a single $^{9}$Be$^{+}$ ion is confined to a
zone of a trap similar to that in Ref.~\cite{roweQIC}.  The ion is
optically pumped to the state $|A\rangle$, and its motion is
Doppler cooled by use of the cycling transition $|A\rangle
\leftrightarrow |2p$~$^{2}P_{3/2}, F' = 3, m_{F'} =
3\rangle$~\cite{sideband_cool}.  We detect the state of the
$^{9}$Be$^{+}$ ion through state-dependent resonance
fluorescence~\cite{sideband_cool}.  That is, with light tuned to
the cycling transition, the state $|A\rangle$ fluoresces strongly,
whereas the other states do not.  Using coherent rotations
described below, we measure the $|$$\downarrow\rangle$,
$|$$\uparrow\rangle$ ``qubit'' level populations by mapping the
states $|$$\uparrow$$\rangle$ and $|$$\downarrow$$\rangle$ to
$|A\rangle$ and $|$$\uparrow$$\rangle$ respectively and measuring
the state $|A\rangle$.

Coherent rotations between states
$|A\rangle$~$\leftrightarrow$~$|$$\uparrow$$\rangle$ and
$|$$\uparrow$$\rangle$~$\leftrightarrow$~$|$$\downarrow$$\rangle$
have the form (in the Bloch sphere representation)

\begin{equation}
R(\theta,\phi)= \cos\frac{\theta}{2}\ I - i
\sin\frac{\theta}{2}\cos\phi\ \sigma_x - i
\sin\frac{\theta}{2}\sin\phi\ \sigma_y, \label{rotation}
\end{equation}

\noindent where $I$ is the identity matrix, $\sigma_{i}$ are Pauli
operators, $\theta$ is the rotation angle, and $\phi$ is the angle
from the x-axis to the rotation axis (in the x-y plane). These
rotations are driven by two-photon stimulated Raman transitions
using focused laser beams~\cite{bible,sideband_cool}. We modulate
one polarization component of a single laser beam with an
electro-optic modulator. This technique simplifies the
stabilization of differential optical path length fluctuations
between the two Raman beams (generated by the two polarizations),
similar to Ref.~\cite{lee03}. The difference in optical path is
due solely to the static birefringence of the modulator's
electro-optic crystal. We stabilize fluctuations in the
birefringence by measuring the retardation with an optical phase
detector similar to that in Ref.~\cite{hanche_cavity} and feeding
back on the temperature of an additional birefringent crystal in
the beam path. Small offsets of the retardation from its optimal
value of $\lambda$/4 only reduce the magnitude of the Rabi
frequency to second order and do not alter the phase of subsequent
rotations.

\begin{figure}
\includegraphics{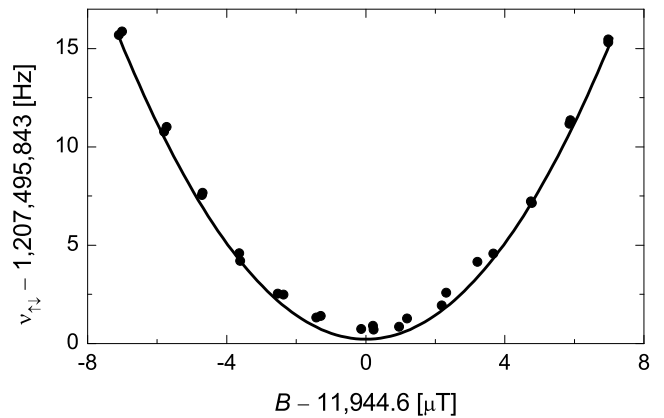} \caption{\label{parabola}Frequency of the
field-independent transition
$|$$\downarrow$$\rangle$~$\leftrightarrow$~$|$$\uparrow$$\rangle$
as a function of magnetic field. Circles are measured data points;
the solid curve is a theoretical prediction.  The statistical
uncertainty of each datum is $\Delta$$B$~$\lesssim$~3~nT and
$\Delta$$\nu_{\uparrow\downarrow}$~$\lesssim$~0.3~Hz.}
\end{figure}

To characterize the field-independent transition, we perform
Ramsey spectroscopy~\cite{mol_beams} on the two transitions $|A
\rangle$~$\leftrightarrow$~$|$$\uparrow$$\rangle$ and
$|$$\downarrow$$\rangle$~$\leftrightarrow$~$|$$\uparrow$$\rangle$
for different magnetic fields (Fig.~\ref{parabola}). The magnetic
field is determined from the $\nu_{\uparrow A}$ measurement. By
measuring $\nu_{\uparrow\downarrow}$ at $B=B_0$ for different RF
trapping strengths and extrapolating to zero, we can determine the
corresponding AC Zeeman shift produced by the trap's RF currents.
This shift [1.81(2)~Hz] was removed from the data in
Fig.~\ref{parabola}. The calculated solid curve in
Fig.~\ref{parabola} is derived from data in
Refs.~\cite{Be_gJ,bollinger}.

\begin{figure}
\includegraphics{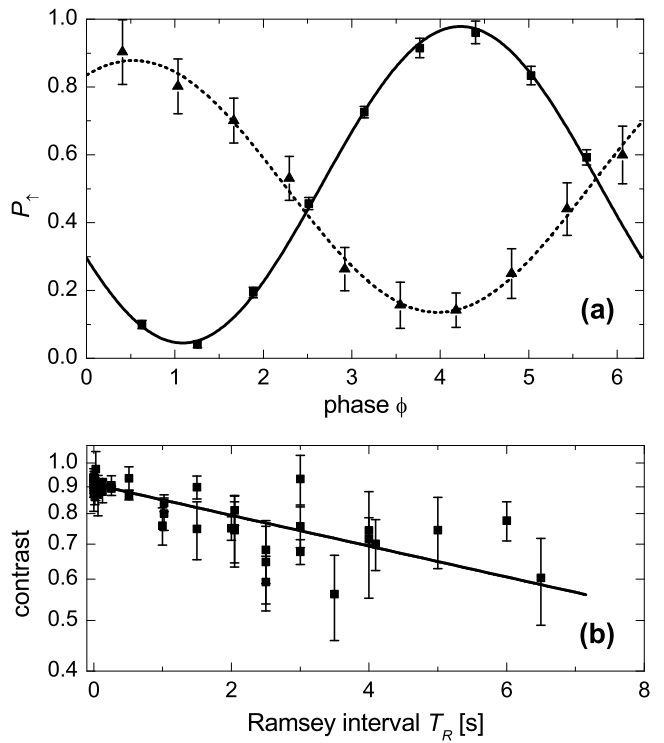} \caption{\label{phase_contrast}\textbf{(a)}
Ramsey data at $T_{R}$~=~4~ms (squares) and 4~s (triangles). The
y-axis represents the probability of measuring the state
$|$$\uparrow$$\rangle$. The number of measurements per phase point
is 1000 and 100 for the 4~ms and 4~s phase scans, respectively.
The contrast $b$ for the 4~ms data is 0.933~$\pm$~0.014 and for
the 4~s data is 0.742~$\pm$~0.043.  The $\phi_D \simeq$1~rad phase
shift in the 4~ms data is due to detuning the local oscillator by
the differential Stark shift ($\sim$4.2~kHz) such that the Ramsey
$\pi$/2-pulses are resonant. \textbf{(b)} Contrast vs. Ramsey
interval $T_{R}$.  Each datum represents the fitted contrast $b$
for a phase scan with Ramsey interval $T_{R}$.  The solid curve is
a weighted least-squares fit to the data.}
\end{figure}

We measure the qubit coherence time by tuning the magnetic field
to the minimum of Fig.~\ref{parabola} and performing Ramsey
spectroscopy on the
$|$$\downarrow$$\rangle$~$\leftrightarrow$~$|$$\uparrow$$\rangle$
transition for different Ramsey intervals $T_{R}$.  The
$^{9}$Be$^{+}$ ion is first Doppler cooled and prepared in the
state $|$$\uparrow\rangle$.  We then apply the rotation
$R(\frac{\pi}{2},0)$, creating the superposition state
$|\Psi_1\rangle = \frac{1}{\sqrt{2}}(|$$\uparrow$$\rangle - i
|$$\downarrow$$\rangle)$ and wait for the Ramsey interval $T_{R}$
during which the state evolves to $|\Psi_2\rangle =
\frac{1}{\sqrt{2}}(e^{i\phi_D}|$$\uparrow\rangle - i |$$\downarrow
\rangle )$.  The phase $\phi_D$ is given by the integrated
detuning of the local oscillator frequency (which modulates the
Raman laser frequency) with respect to the qubit transition
frequency over the Ramsey interval $T_R$. A second rotation
$R(\frac{\pi}{2},\phi)$ is then applied to our qubit with $\phi$
variable. Repeating the experiment many times and performing a
projective measurement of the state $|$$\uparrow$$\rangle$ as
described above yields

\begin{equation}
P_{\uparrow} = \frac{1}{2} (1 - \cos{(\phi_D+\phi)}),
\label{detect}
\end{equation}

\noindent the probability of measuring the state
$\vert$$\uparrow$$\rangle$.  The measurement sequence is repeated
for different phases~$\phi$, and the detected
probability~$P_{\uparrow}$ is fit to the function $f = a -
\frac{b}{2} \cos{(d \phi + \phi_{D})}$.  The fit parameter~$d$
allows for magnetic-field drift in time as successive phase points
are recorded sequentially; $d$ is close to unity for all scans in
this data set. Phase scans for $T_{R}$~=~4~ms and 4~s are shown in
Fig.~\ref{phase_contrast}a. Any phase fluctuation of the qubit
state with respect to our local oscillator (derived from a
Hydrogen maser source with negligible phase uncertainty compared
to that of the qubit state) during the Ramsey interval $T_{R}$
will reduce the contrast~$b$. The contrast~$b$ for different $T_R$
is fit to the exponential $b(T_{R}) = b_0 e^{-T_{R}/\tau}$
(Fig.~\ref{phase_contrast}b). We find the coherence time for our
field-independent qubit to be $\tau = 14.7 \pm 1.6$~s. The qubit
coherence time is limited by slow drift of the magnetic field over
the measurement time scale of a single point. For the $T_R=4$~s
data in Fig.~\ref{phase_contrast}a, this time scale is 400~s. In
principle, if the magnetic-field drift is small for the period of
a single measurement, we can interrupt data collection to measure
(via $\nu_{\uparrow A}$) and correct for magnetic-field deviations
from $B_0$.

Logical qubits of the decoherence-free subspace (DFS)~\cite{DFS}
comprising two entangled physical qubits in the form of Bell
states,

\begin{equation}
|\Psi_{\pm}\rangle=\frac{1}{\sqrt{2}}(|01\rangle \pm |10\rangle),
\label{singletTriplet}
\end{equation}

\noindent are also immune to fluctuations in magnetic fields.  Any
phase acquired due to a (uniform) fluctuation of magnetic field by
one state of the superposition is acquired equally by the other
state of the superposition.  In the experiment described below
(performed in a separate but similar trap), the physical qubit
states $|0\rangle$ and $|1\rangle$ are the
magnetic-field-sensitive hyperfine states $|F=1, m_F=-1\rangle$
and $|F=2, m_F=-2\rangle$ respectively at a field $B \simeq
0.0013$~T. Using the technique of Ref.~\cite{roos04} as described
below, we demonstrate that entanglement is long-lived.

Even though the states $|\Psi_{\pm}\rangle$ are immune to uniform
time-varying magnetic fields, they are \textit{not} protected from
noise in a magnetic-field difference between the locations of the
two ions. Such a gradient can cause the states $|01\rangle$ and
$|10\rangle$ to acquire phase at a differential rate
$\Delta\phi(t)$ due to the different local magnetic fields.  This
results in a coherent oscillation between $|\Psi_+\rangle$ and
$|\Psi_{-}\rangle$ according to

\begin{equation}
|\psi(t)\rangle = \cos\Bigl[\frac{\Delta\phi(t)}{2}
\Bigr]|\Psi_{+}\rangle + i \sin\Bigl[\frac{\Delta\phi(t)}{2}
\Bigr]|\Psi_{-}\rangle. \label{oscillation}
\end{equation}

Before each experiment we perform Doppler cooling,
resolved-sideband cooling, and optical pumping to bring the two
ions to the vibrational ground state in the trap with internal
state $|11\rangle$~\cite{kingcooling}.  As described
in~\cite{didi_gate}, we prepare the maximally entangled state

\begin{equation}
|\Phi_{-i}\rangle=\frac{1}{\sqrt{2}}(|00\rangle - i |11\rangle).
\label{didigatestate}
\end{equation}

\noindent Following this step, we apply a rotation
$R(\frac{\pi}{2},-\frac{\pi}{4})$ to both ions to create the
state, $|\Psi_+\rangle$.

After preparation of the $|\Psi_+\rangle$ state, we wait for a
delay $t_D$ and then apply a final rotation $R(\frac{\pi}{2},0)$
to both qubits.  This transforms $|\Psi_+\rangle$ into the Bell
state $|\Phi_{+}\rangle=\frac{1}{\sqrt{2}}(|00\rangle +
|11\rangle)$, but does not affect the singlet state
$|\Psi_-\rangle$ as it is invariant under collective rotations. We
detect both ions simultaneously; from the fluorescence count
distributions, we can determine the parity of the final state
\cite{rowe01} and therefore the probabilities of $|\Psi_+\rangle$
and $|\Psi_-\rangle$ in Eq.~\ref{oscillation} as a function of
$t_D$.

\begin{figure}
\includegraphics{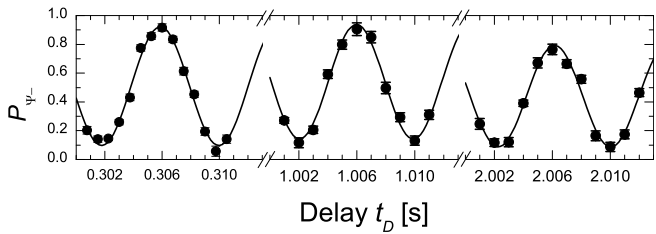}
\caption{\label{dfsflop}Coherent oscillation between
$|\Psi_+\rangle$ and $|\Psi_-\rangle$ states as a function of
delay $t_D$. $P_{\Psi_-}$ represents the probability of measuring
$|\Psi_-\rangle$. The line is a sinusoidal fit to the data. Data
are shown after delays of 300~ms, 1~s, and 2~s.}
\end{figure}

Figure~\ref{dfsflop} displays data for the coherent oscillation
around three different delays $t_D$.  For these data, the
magnetic-field gradient induces an oscillation frequency of
approximately 125~Hz.
From the decay of the $|\Psi_+\rangle$, $|\Psi_-\rangle$
oscillations with delay $t_D$ we extract a Bell-state lifetime of
$7.3$~$\pm$~1.6~s, assuming exponential decay. The measured
entanglement lifetime was limited by fluctuations in the
magnetic-field gradient.

Combining field-independent qubits and DFS states, we could create
memories even more robust than the DFS Bell states demonstrated
here and in Ref.~\cite{roos04}, as different local magnetic fields
will induce very small frequency shifts of the qubit transition.
The other two states in the Bell basis
$\frac{1}{\sqrt{2}}(|00\rangle \pm |11\rangle)$ will also benefit
from reduced decoherence due to magnetic field noise.

In summary, we have shown how field-independent qubits can serve
as good memory elements in a trapped-ion-based quantum information
processor.  Decoherence-free-subspace (DFS) qubits as demonstrated
here and in~\cite{DFS,roos04} can also be used as good memory
elements, with the additional overhead of encoding into the DFS
states.  Combining both techniques should lead to memory elements
with extremely long coherence times.  One disadvantage of
field-independent qubits is that gates relying on differential
Stark shifts between the qubit states \cite{didi_gate} will cease
to work when the qubit transition frequency is small compared to
the Raman beam detuning from the excited states. To overcome this
limitation, we can momentarily change the qubit states, perform
the gate, and transform back to the original qubit basis. If the
ambient magnetic fields fluctuate on time scales much longer than
the duration of these three steps, accumulated phase errors should
be negligible. Alternatively, we could apply a gate in which both
bits are simultaneously
flipped~\cite{sackett00,molmer99,haljan05}.

The demonstration of robust qubit memories and long-lived
entanglement in trapped atomic ion systems satisfies one of the
requirements necessary for large scale quantum information
processing.  Assuming exponential decay, the probability of memory
error is $1.4 \times 10^{-5}$ for current detection durations of
200~$\mu$s, which is below the fault-tolerant thresholds set by
Steane~\cite{steane03} and Knill~\cite{knill05}.  This, in
combination with the ability to reduce spontaneous emission errors
during laser excitation~\cite{ozeri05}, makes atomic ion systems
promising candidates for fault-tolerant QIP.

The authors thank Jeroen Koelemeij and Signe Seidelin for helpful
comments on the manuscript.  This work was supported by the US
National Security Agency (NSA) and the Advanced Research and
Development Activity (ARDA) under contract number MOD-7171.05.
This manuscript is a publication of NIST and is not subject to U.
S. copyright.

\end{document}